\documentclass[12pt]{article}
\usepackage{epsf}
\usepackage{amsmath}

\usepackage{graphics}
\usepackage{cite}

\begin{document}
\begin{titlepage}

\hfill April 2023

\begin{center}

{\bf \LARGE Time Dependence of Charged Dark Matter}\\
\vspace{2.5cm}
{\bf Paul H. Frampton}\footnote{paul.h.frampton@gmail.com}\\
\vspace{0.5cm}
{\it Dipartimento di Matematica e Fisica "Ennio De Giorgi",\\ 
Universit\`{a} del Salento and INFN-Lecce,\\ Via Arnesano, 73100 Lecce, Italy.
}

\vspace{1.0in}

\begin{abstract}
\noindent
We investigate a model of the universe where dark energy is replaced by
electrically-charged extremely-massive dark matter. This was originally
described only for the present cosmological time. The time dependence of the
charged dark matter is different from that for dark energy and in the
future the expansion will no longer accelerate and the scale
factor $a(t)$ will revert to a matter-dominated behaviour
$a(t) \sim t^{\frac{2}{3}}$. The consequences for the introverse and
extroverse are discussed. Unlike in a
$\Lambda CDM$ model, many other galaxies will always remain visible.
\end{abstract} 
\end{center}

\end{titlepage}

\noindent
\section{Introduction}

\noindent
A novel cosmological model was recently suggested \cite{FramptonPLB,FramptonMPLA} in which dark energy is
replaced by charged dark matter in the form of cPEMBHs or charged
Primordial Extremely Massive Black Holes. That discussion focused on the present cosmological
time $t=t_0\simeq 13.8$ Gy and already provided some counterintuitive ideas such as that at the largest cosmological distances, {\it e.g.} greater than 1 Gpc, the dominant force
is electromagnetism rather than gravitation. 
\bigskip

\noindent
The production mechanism for PBHs in general is not well understood, and for the
cPEMBHs we shall make the simplifying assumption that they are first formed
when the accelerated expansion begins at $t=t_{DE}\sim 9.8$ Gy, as in Table 1
in \cite{FramptonIJMPA}. For the expansion before $t_{DE}$ we shall assume
that the $\Lambda CDM$ model is approximately accurate.

\bigskip

\noindent
The subsequent expansion in the charged dark matter cPEMBH model will in the future
depart markedly from the $\Lambda CDM$ case. We can regard this as advantageous
because the future fate of the universe in the conventional picture does have certain
distasteful features in terms of the extroverse, as we bieifly
review.

\bigskip

\noindent
In the $\Lambda CDM$  model the introverse, or what is also called the visible 
universe, coincides with the extroverse at $t=t_{DE} \sim 9.8$ Gy with the common
radius
\begin{equation}
R_{EV}(t_{DE}) = R_{IV}(t_{DE}) =  39 Gly
\label{tDE}
\end{equation}
according to Table 3 in \cite{FramptonIJMPA}).

\bigskip

\noindent
The introverse expansion is limited by the speed of light and its radius increases
from Eq. (\ref{tDE}) to 44 Gly at the present time $t=t_0$ and asymptotes to
\begin{equation}
R_{IV} (t \rightarrow \infty) = 58 Gly
\label{RIVasymp}
\end{equation}

\bigskip

\noindent
The extroverse expansion is exponential and superluminal. Its radius increases
from its\\value 39 Gly in Eq. (\ref{tDE}) to 52 Gly at the present time $t=t_0$ and grows without limit
so that after a trillion years it attains the extremely large value
\begin{equation}
R_{EV} (t  = 1 Ty) = 9.7 \times 10^{32} Gly.
\label{REVtrillion}
\end{equation}

\bigskip

\noindent
This future for the $\Lambda CDM$ scenario seems distasteful because the
introverse becomes of ever decreasing, and eventually vanishing, significance,
relative to the extroverse.

\newpage

\section{The future of charged dark matter}

\noindent
A possible formation mechanism of cPEMBHs was provided
in \cite{Araya2022} where their common sign of electric charge, negative,
arises from preferential accretion of electrons relative to protons. This
formation mechanism is not well understood\footnote{The cPEMBHs discussed in \cite{Araya2022}, while inspirational,
do not experience long-range electromagnetic forces \cite{Araya2023}.
The cPEMBHs in the theory proposed by \cite{FramptonIJMPA}
therefore require a modified production mechanism.}
\footnote{A possible test for the presence of cPEMBHs was proposed
in \cite{Wagner}.}
\footnote{Electrically neutral PEMBHS were first considered, with a different acronym SLABs,  in \cite{Carr2021}.}
so to create a cosmological model we shall for simplicity assume that
the cPEMBHs are all formed between $t=t_{DE} \sim 9.8$ Gy and
$t_0 \sim 13.8$Gy. As discussed in \cite{FramptonMPLA}, the
Friedmann equation ignoring radiation, during this time window, is
\begin{equation}
\left( \frac{\dot{a}}{a} \right)^2= \frac{\Lambda}{3} + \frac {8\pi G}{3} \rho_{matter}
\label{Friedmann}
\end{equation}
where $\Lambda$ is the cosmological constant generated by the Coulomb
repulsion between the cPEMBHs.
From Eq.(\ref{Friedmann}), with $a(t_0) = 1$ and constant $\Lambda$, we
would predict that
\begin{equation}
a(t \rightarrow \infty) \sim exp \left( \sqrt{ \frac{\Lambda}{3}} (t-t_0) \right)
\label{exponential}
\end{equation}

\bigskip

\noindent
However, in the case of charged dark matter, with no dark energy, we must
re-write \\Eq, ({\ref{Friedmann}) as
\begin{equation}
\left( \frac{\dot{a}}{a} \right)^2= \frac {8\pi G}{3} \rho_{cPEMBHs}+ \frac {8\pi G}{3} \rho_{matter}
\label{FriedmannPrime}
\end{equation}
in which
\begin{equation}
\rho_{matter} (t)  = \frac{\rho_{matter} (t_0)}{a(t)^3}
\label{rhomatter}
\end{equation}
where matter includes both normal matter and the uncharged dark matter.

\bigskip

\noindent
Of special interest in the present discussion is the expected future behaviour
of the charged dark matter
\begin{equation}
\rho_{cPEMBHs} (t) = \frac{\rho_{cPEMBHs} (t_0)}{a(t)^3}
\label{rhocPEMBHs}
\end{equation}
so that comparison of Eq.(\ref{Friedmann}) and Eq.(\ref{FriedmannPrime}) suggests
that the cosmological constant is predicted to decrease from its present value.
More specifically, we find
that asymptotically  the scale factor will behave as if matter-dominated
and the cosmological constant will decrease at large future times as a power
\begin{equation}
a(t\rightarrow \infty) \sim t^{\frac{2}{3}} ~~~~~~ \Lambda(t \rightarrow \infty) \sim t^{-2}.
\label{scale}
\end{equation} 

\bigskip

\noindent
so that a trillion years in the future $\Lambda(t)$ will have decreased
by some four orders of magnitude relative to $\Lambda(t_0)$.

\newpage

\section{Discussion}

\noindent
According to the $\Lambda CDM$ model, we live at a special time in
cosmic history because of the density coincidence between dark matter
and dark energy.
In the case of charged dark matter replacing dark energy, the present
era is even more  special because the striking accelerated expansion,
discovered in 1998, is a temporary phenomenon centred around the present time.
Acceleration 
began about 4 Gy ago at $t_{DE}= 9.8Gy = t_0-4 Gy$.
This unexpected behaviour will disappear in a few more billion
years. The value of the cosmological constant is predicted to fall
like $a(t)^{-2}$ so that, when $t \sim \sqrt{2} t_0 \sim 19.5 Gy
\sim t_0 +4.7 Gy$, the value of
$\Lambda(t)$ will be one half of its present value, $\Lambda(t_0)$. 
As discussed in \cite{FramptonMPLA}, the equation of state associated 
with $\Lambda$ is predicted to be
extremely close to $\omega = -1$, so close that measuring the
difference seems impracticable.

\bigskip

\noindent
Let us  discuss the future time evolution of the introverse and
extroverse in the case of charged dark matter.  For the introverse,
nothing changes from the $\Lambda CDM$ case discussed in Table 3
of \cite{FramptonIJMPA}. After a trillion years, the introverse radius
will be at its asymptotic value $R_{IV}=58 Gly$, as stated in Eq.(\ref{RIVasymp}).
By contrast, the future for the extroverse
is very different for charged dark matter. 
WIth the growth $a(t) \propto t^{\frac{2}{3}}$ we find
that the radius of the extroverse at $t=1$ Ty is 
\begin{equation}
R_{EV}(t=1Ty) \sim 900 Gly
\label{REVnew}
\end{equation}

\bigskip

\noindent
to be compared with the corresponding huge value
$9.7 \times 10^{32}$ Gly predicted by the $\Lambda CDM$ model, quoted in
Eq.(\ref{REVtrillion}) above. Eq.(\ref{REVnew}) means that if there still exist humans
in the Solar System, or at least in the Milky Way, their view
of the distant universe will include many billions of galaxies.

\bigskip

\noindent
In the $\Lambda CDM$ case, a hypothetical observational cosmologist, one trillion years
in the future,
could observe only the Milky Way and objects such as the Magellanic 
Clouds which are gravitationally bound to it, so that cosmology could become
an extinct science.
In the case of charged dark matter, for comparison, the time dependence
will allow about 180 billion out of a present trillion galaxies to
remain observable at $t=1Ty$ so that the view of the 
universe at that distant future time will look quite similar to 
the view at the present.
.
\bigskip

\noindent
The distinct physics advantage of charged dark matter is that it avoids the 
idea of an unknown repulsive gravity inherent in dark energy. Electromagnetism
provides the only known long-range repulsion so it is more attractive
to adopt it as the explanation for the accelerating universe.
A second advantage of charged dark matter 
is that it provides a conducive environment for cosmology, a trillion years in the
future.


\newpage

\begin{thebibliography}{999}

\bibitem{FramptonPLB}
P.H. Frampton,\\
{\it Electromagnetic Accelerating Universe}.\\
Phys. Lett. {\bf B835,} 137480 (2022).\\
{\tt arXiv:2210.10632}.

\bibitem{FramptonMPLA}
P.H. Frampton,\\
{\it  A Model of Dark Matter and Energy}.\\
Mod. Phys. Lett. {\bf A,}  (2023 in press).\\
{\tt arXiv:2301.10719}.

\bibitem{FramptonIJMPA}
P.H. Frampton,\\
{\it Cyclic Entropy:An Alternative to Inflationary Cosmology}.\\
Int. J.Mod. Phys. {\bf A30,} 1550129 (2015).\\
{\tt arXiv:1501.03054}.

\bibitem{Araya2022}
I.J. Araya, N.D. Padilla, M.E. Rubio, J. Sureda, J. Magana and L. Osorio,\\
{\it Dark Matter from Primordial Black Holes Would Hold Charge}.\\
JCAP {\bf 02:}030 (2023).\\
{\tt arXiv:2207.05829}.

\bibitem{Araya2023}
I.J. Araya, private communication (2023).

\bibitem{Wagner}
J. Wagner,\\
{\it Observables for Moving, Extremely Charged Primordial Extremely \\
Massive Black Holes}.\\
{\tt arXiv:2301.13210}. (2023).

\bibitem{Carr2021}
B. Carr, F. Kuhnel and L. Visinelli,\\
{\it Constraints on Stupendously Large Black Holes}.\\
MNRAS {\bf 501,} 2029 (2021).\\
{\tt arXiv:2008.08077}.


\end{thebibliography}
\end{document}